\documentclass[aps,prb,twocolumn,preprintnumbers,amsmath,amssymb,superscriptaddress,nopacs]{revtex4-1}

\usepackage{graphicx}
\usepackage{bm}
\usepackage[usenames,dvipsnames]{xcolor}
\usepackage{hyperref}
\usepackage{pstricks}
\usepackage[tight]{subfigure}
\usepackage{verbatim}
\usepackage{units}
\usepackage{multirow}
\usepackage{enumitem}
\usepackage{leftidx}
\usepackage{lipsum}

\usepackage{t1enc}
\usepackage[polish,english]{babel}
\usepackage[cp1250]{inputenc}
\usepackage{polski}
\usepackage[T1]{fontenc}
\renewcommand{\vec}[1]{\mathbf{#1}}

\newcommand{\vecG}[1]{\bm{#1}}

\newcommand{\exch}{\Delta\varepsilon_{\rm exch}}

\newcommand{\ketin}[2]{|#1\rangle_{\!\raisebox{-3pt}{\scalebox{0.6}{\textrm{#2}}}}}
\newcommand{\brainline}[2]{\leftidx{_{\raisebox{-1pt}{\scalebox{0.6}{\textrm{#2}}}}\!}{\langle #1|}}
\newcommand{\ketinline}[2]{|#1\rangle_{\!\raisebox{-2pt}{\scalebox{0.6}{\textrm{#2}}}}}

\usepackage[normalem]{ulem}

\begin{document}

\title{Transverse anisotropy effects on spin-resolved transport through large-spin molecules}

\author{Maciej Misiorny}
 \email{misiorny@amu.edu.pl}
\affiliation{Peter Gr{\"u}nberg Institut, Forschungszentrum J{\"u}lich, 52425 J{\"u}lich,  Germany}
\affiliation{JARA\,--\,Fundamentals of Future Information Technology, 52425 J{\"u}lich,  Germany}
\affiliation{Faculty of Physics, Adam Mickiewicz University, 61-614 Pozna\'{n}, Poland}

\author{Ireneusz Weymann}
\affiliation{Faculty of Physics, Adam Mickiewicz University,
61-614 Pozna\'{n}, Poland}

\date{\today}

\begin{abstract}
The transport properties of a large-spin molecule
strongly coupled to ferromagnetic leads in the presence
of transverse magnetic anisotropy are studied theoretically.
The relevant spectral functions, linear-response
conductance and the tunnel magnetoresistance
are calculated by means of the numerical renormalization group method.
We study the dependence of transport characteristics on
orbital level position, uniaxial and transverse anisotropies,
external magnetic field and  temperature.
It is shown that while uniaxial magnetic anisotropy leads to the suppression
of the Kondo effect, finite transverse anisotropy can restore the
Kondo resonance. The effect of Kondo peak restoration
strongly depends on the magnetic configuration of the device and
leads to nontrivial behavior of the tunnel magnetoresistance.
We show that the temperature dependence of the conductance at
points where the restoration of the Kondo effect occurs
is universal and shows a scaling typical for usual spin-one-half Kondo effect.
\end{abstract}

\pacs{72.10.Fk,72.25.-b,73.23.-b,75.50.Xx}



\maketitle


\section{Introduction}

Knowledge of transport properties of individual large-spin ($S> 1/2$) atoms~\cite{Otte_NaturePhys.4/2008,Khajetoorians_Science332/2011,Loth_Science335/2012} or single-molecule magnets (SMMs)~\cite{Gatteschi_book,Heersche_Phys.Rev.Lett.96/2006,Zyazin_NanoLett.10/2010,Burzuri_Phys.Rev.Lett.109/2012,Parks_Science328/2010,Vincent_Nature488/2012} that exhibit magnetic anisotropy is of key importance from the point of view of information processing technologies.~\cite{Mannini_NatureMater.8/2009,Khajetoorians_Science332/2011} The ultimate aim is to incorporate such objects as functional elements of spintronic devices, with the objective of employing spin-polarized currents to control the magnetic state of the system. In particular, for an atom/molecule with the predominant \emph{`easy-axis'} uniaxial magnetic anisotropy this allows for switching the system's spin between two metastable states.~\cite{Misiorny_Phys.Rev.B75/2007,Timm_Phys.Rev.B73/2006,Delgado_Phys.Rev.Lett.104/2010,
Fransson_Phys.Rev.B81/2010,Loth_NaturePhys.6/2010} However, apart from the \emph{uniaxial} magnetic anisotropy underlying the magnetic bistability, adatoms and SMMs usually  possess also the \emph{transverse} component of the anisotropy.~\cite{Gatteschi_book} If the latter component is sufficiently large, not only may it impede the spin switching process~\cite{Misiorny_Phys.Rev.Lett.111/2013}, but also it leads to additional quantum effects, such as oscillations due to the geometric Berry phase~\cite{Wernsdorfer_Science284/1999} or quantum tunneling of magnetization (QTM),~\cite{Sessoli_Nature365/1993,Thomas_Nature383/1996,Mannini_Nature468/2010,Gatteschi_Angew.Chem.Int.Ed.42/2003} which can manifest in transport characteristics.~\cite{Kim_Phys.Rev.Lett.92/2004,Romeike_Phys.Rev.Lett.96/2006a,Gonzalez_Phys.Rev.Lett.98/2007}

Equally interesting is the situation of the strong-coupling regime,
where the Kondo correlations emerge, so that anomalous signatures
in transport become apparent for temperatures lower than the Kondo temperature~$T_\text{K}$.
\cite{Kondo_Prog.Theor.Phys32/1964,Hewson_book}
Notably, for spin-one-half ($S=1/2$) systems the linear-response conductance can then achieve the unitary limit of $2e^2/h$.
\cite{Goldhaber_Nature391/98,Cronenwett_Science281/1998}
In the case of $S>1/2$, depending on the number of screening
channels, one can observe more exotic types of the Kondo effect,
such as, e.g., the underscreened Kondo effect, which emerges
when the number of screening channels is smaller than $2S$.
\cite{Nozieres_J.Phys.(France)41/1980,LeHur_Phys.Rev.B56/1997}
Such situation occurs in molecular or in left-right asymmetric junctions.
\cite{Roch_Phys.Rev.Lett.103/2009,Parks_Science328/2010}
In large-spin systems a key factor determining whether the Kondo effect will occur or not
is actually the magnetic anisotropy.
\cite{Zitko_Phys.Rev.B78/2008,Zitko_J.Phys.:Condens.Matter22/2010,Misiorny_Phys.Rev.B86/2012_UK}
For a sole \emph{uniaxial} component of magnetic anisotropy,
the effect was observed only if the \emph{planar} state was preferred,
\cite{Otte_NaturePhys.4/2008,Parks_Science328/2010}
and expected to be inhibited otherwise
\cite{Misiorny_Phys.Rev.Lett.106/2011,Misiorny_Phys.Rev.B84/2011,Elste_Phys.Rev.B81/2010}
---single-electron spin-exchange processes within the ground state doublet
involving the \emph{axial} states are forbidden for $S>1/2$.
Interestingly, in the latter case the Kondo effect can be in principle restored
if one allows for mixing of the two axial states, which can be accomplished
by introduction of the transverse component of magnetic anisotropy.
\cite{Romeike_Phys.Rev.Lett.96/2006,Romeike_Phys.Rev.Lett.97/2006,
Leuenberger_Phys.Rev.Lett.97/2006}

Although the role of the transverse magnetic anisotropy in the
formation of the Kondo effect has been studied extensively for normal electrodes,~\cite{Romeike_Phys.Rev.Lett.96/2006,Romeike_Phys.Rev.Lett.97/2006,
Leuenberger_Phys.Rev.Lett.97/2006,Wegewijs_NewJ.Phys.9/2007,Gonzalez_Phys.Rev.B78/2008,
Roosen_Phys.Rev.Lett.100/2008,Zitko_Phys.Rev.B78/2008,Zitko_NewJ.Phys.11/2009,
Zitko_NewJ.Phys.12/2010,Romero_arXiv:1404.3755}
not much is known about spin-polarized transport in such a case.
For this reason, in the present paper we address the problem of spin-resolved transport
through large-spin nanostructures in the presence of transverse magnetic anisotropy
focusing on the Kondo regime.
The presence of ferromagnetic electrodes results in exchange fields,
which can lead to the spin-splitting of the levels of the nanostructure.
\cite{Martinek_Phys.Rev.Lett.91/2003_127203,Martinek_Phys.Rev.Lett.91/2003_247202,
Choi_Phys.Rev.Lett.92/2004,Pasupathy_Science306/2004,Hauptmann_NaturePhys.4/2008,
Gaass_Phys.Rev.Lett.107/2011,Lim_Phys.Rev.B88/2013,Misiorny_NaturePhys.9/2013}
The exchange fields are thus another relevant energy scale in the problem,
which determines the occurrence of the Kondo effect
and thus conditions the transport properties of the system.
In order to reliably analyze the interplay of the effects due to the exchange fields
and magnetic anisotropy on our large-spin nanostructure,
we employ the numerical renormalization group (NRG) method.
\cite{Wilson_Rev.Mod.Phys.47/1975}
This method is known as one of the most accurate methods in studying the transport
properties of various quantum impurity models.
\cite{Bulla_Rev.Mod.Phys.80/2008}

The paper is organized as follows:
In Sec. II we describe the model Hamiltonian and method
used to calculate the transport properties.
Section III contains numerical results and their discussion.
First, the ground state properties of the molecule
are discussed (Sec. III.A), then the behavior
of the relevant spectral functions is analyzed (Sec. III.B).
The level and temperature dependence of the linear conductance and TMR
is presented in Sec. III.C, while in Secs. III.D and III.E
we analyze how transport properties depend on
the anisotropy constants and on transverse magnetic field applied to the system.
At the end of Sec. III we also discuss the universal scaling
of the linear conductance as a function of temperature.
Finally, the paper is concluded in Sec. IV.

\section{Theoretical description}
\subsection{Model}

\begin{figure}
\centering
\includegraphics[width=0.65\columnwidth]{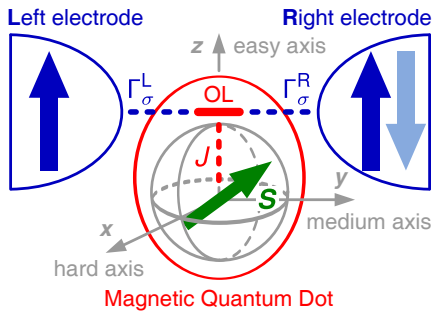}
\caption{(color online)
Schematic of a magnetic quantum dot.
It consists of a single conducting orbital level (OL)
tunnel-coupled to two ferromagnetic electrodes,
with coupling strengths $\Gamma^{\rm L}_\sigma$ and  $\Gamma^{\rm R}_\sigma$,
and exchange-coupled ($J$) to a magnetic core of large spin $S>1/2$
exhibiting both uniaxial and transverse magnetic anisotropies.
The easy axis is oriented along the magnetization of the leads, which
can form either parallel or antiparallel configuration.
}
\label{Fig:1}
\end{figure}

In order to grasp the essential features of a nanoscopic system exhibiting magnetic anisotropy,
we employ a model consisting  of a magnetic core represented by a large spin $S>1/2$,
which is exchange coupled with strength $J$ to a single conducting
orbital level (OL), see Fig.~\ref{Fig:1}. Such generic model of the molecule will henceforward
be referred to as \emph{magnetic quantum dot} (MQD),
and it can be characterized by the Hamiltonian of the form
    \begin{equation}\label{Eq:Ham_MQD}
    \hat{\mathcal{H}}_\textrm{MQD}
    =
     \hat{\mathcal{H}}_\textrm{OL}
    + \hat{\mathcal{H}}_\textrm{S}
    -J \hat{\vec{s}}\cdot \hat{\vec{S}}
    +\vec{B}\cdot( \hat{\vec{s}}+ \hat{\vec{S}}).
    \end{equation}
In the above, the first term describes the orbital level,
    \begin{equation}
    \hat{\mathcal{H}}_\textrm{OL}
    =
    \varepsilon\sum_\sigma  \hat{n}_\sigma + U  \hat{n}_\uparrow  \hat{n}_\downarrow,
    \end{equation}
where $ \hat{n}_\sigma= \hat{c}_\sigma^\dagger  \hat{c}_\sigma^{}$ denotes the occupation operator and $ \hat{c}_\sigma^\dagger ( \hat{c}_\sigma^{})$ stands for the operator creating (annihilating) a spin-$\sigma$ electron of energy $\varepsilon$ in the OL, while $U$ accounts for the Coulomb energy of two electrons of opposite spins dwelling in the orbital. Furthermore, we assume that only the core spin is subject to magnetic anisotropy, and thus, at sufficiently low temperatures, its magnetic properties can be captured by the \emph{giant-spin Hamiltonian}~\cite{Gatteschi_book}
    \begin{equation}\label{Eq:Ham_S}
     \hat{\mathcal{H}}_\textrm{S}
    =
    -D \hat{S}_z^2
    +
    E\big( \hat{S}_x^2- \hat{S}_y^2\big).
    \end{equation}
Here, the first/second term stands for the \emph{uniaxial}/\emph{trans\-verse} magnetic anisotropy, with $D$ and $E$ being the relevant anisotropy parameters, and $ \hat{S}_j$ ($j=x,y,z$) representing the $j$th component of the MQD's core spin operator $ \hat{\vec{S}}$. Note that the transverse component is commonly expressed in terms of  the spin ladder operators, $ \hat{S}_\pm= \hat{S}_x\pm i \hat{S}_y$, taking thus the form $(E/2)\big( \hat{S}_+^2+ \hat{S}_-^2\big)$. We focus then on the case of a system displaying  magnetic an\-i\-so\-tropy of an \emph{easy-axis} type, that is for  $D\!>\!0$, assuming in addition that the transverse anisotropy constant is positive, $E\!>\!0$, and varies within the range $0\!\leqslant\! E/D\!\leqslant\!1/3$.~\cite{Gatteschi_book}
Setting such constrictions on values of $D$ and $E$ allows for distinguishing the principal axes of the system, see Fig.~\ref{Fig:1}.

The next term of the Hamiltonian~(\ref{Eq:Ham_MQD})  is responsible for the exchange interaction between the spin $ \hat{\vec{S}}$ of the MQD's magnetic core and the spin $\hat{\vec{s}}$ of an electron residing in the OL, with
    $
     \hat{\vec{s}}
    =
    \frac{1}{2}
    \sum_{\sigma\sigma^\prime}
     \hat{c}_\sigma^\dagger
     \hat{\vecG{\sigma}}_{\sigma\sigma^\prime}^{}
      \hat{c}_{\sigma^\prime}^{}
    $
and $ \hat{\vecG{\sigma}}\equiv( \hat{\sigma}^x, \hat{\sigma}^y, \hat{\sigma}^z)$ denoting the Pauli spin operator. Since no restriction is im\-posed on the sign of the parameter~$J$, the interaction can be either \emph{ferromagnetic} (FM for $J>0$) or \emph{antiferromagnetic} (AFM for $J<0$). Finally, the last term of Eq.~(\ref{Eq:Ham_MQD}) accounts for the Zeeman interaction, where $\vec{B}=(B_x,B_y,B_z)$ corresponds to an external magnetic field measured in energy units, i.e. $g\mu_\textrm{B}\equiv1$.

Transport of electrons through the system is assumed to take place only \emph{via} the OL, which is tunnel-coupled to two ferromagnetic metallic electrodes, see Fig.~\ref{Fig:1}. It is worth a note, however,  that although the magnetic core is not tunnel-coupled directly to electrodes, and thus it does not participate actively in transport, it is still affected by their presence due to the exchange interaction with conduction electrons occupying the OL. The $q$th electrode [$q=\textrm{(L)eft}, \textrm{(R)ight}$] is modelled as a reservoir of  noninteracting itinerant electrons, and described by
    \begin{equation}
     \hat{\mathcal{H}}_\textrm{el}^q
    =
    \sum_\sigma\int_{-W}^W\textrm{d}\epsilon\,\epsilon\,
    \hat{a}_{q\sigma}^\dagger(\epsilon)
    \hat{a}_{q\sigma}^{}(\epsilon),
    \end{equation}
where $ \hat{a}_{q\sigma}^\dagger(\epsilon)$ is the relevant operator responsible for creation of a spin-$\sigma$ electron and $W$ denotes the band half-width. In general, the orientation of electrodes' magnetic moments  with respect to each other and the system's principal axes can be arbitrary. For the sake of simplicity, though, at present we limit the discussion to the situation when magnetic moments of electrodes are collinear (parallel or antiparallel), and their orientation also coincides with that of the system's easy axis. In such a case, tunneling of electrons between the MQD and electrodes is characterized by
    \begin{equation}
     \hat{\mathcal{H}}_\textrm{tun}
    =
    \sum_{q\sigma}\sqrt{\frac{\Gamma_\sigma^q}{\pi}}
    \int_{-W}^W\textrm{d}\epsilon
    \Big[
    \hat{a}_{q\sigma}^\dagger(\epsilon)
    \hat{c}_\sigma^{}
    +
    \hat{c}_\sigma^\dagger
    \hat{a}_{q\sigma}^{}(\epsilon)
    \Big],
    \end{equation}
with $\Gamma_\sigma^q$ representing the  strength of spin-dependent tunnel coupling (hybridization) between the OL and the $q$th electrode. Assuming now that both electrodes are made of the same material, described by the spin polarization coefficient $P$, the hybridization can be parameterized as: $\Gamma_{\uparrow(\downarrow)}^L\!=\!(\Gamma/2)(1\pm P)$ and $\Gamma_{\uparrow(\downarrow)}^R\!=\!(\Gamma/2)(1\mp P)$~for~the \emph{antiparallel} magnetic configuration of electrodes, and $\Gamma_{\uparrow(\downarrow)}^L\!=\!(\Gamma/2)(1\pm P)$ and $\Gamma_{\uparrow(\downarrow)}^R\!=\!(\Gamma/2)(1\pm P)$ for the \emph{parallel}~one.

\subsection{Method}
To analyze the influence of transverse magnetic anisotropy on the
linear-response transport properties of a large-spin molecule in the strong tunnel-coupling (Kondo) regime,
we calculate the \emph{linear conductance} $G$ from the
formula~\cite{Meir_Phys.Rev.Lett.68/1992}
    \begin{equation}\label{Eq:G}
    G=\frac{2e^2}{h}\sum_\sigma\frac{2\Gamma_\sigma^L\Gamma_\sigma^R}
    {\Gamma_\sigma^L+\Gamma_\sigma^R}\int \! \textrm{d}\omega
    \left(\! -\frac{\partial f(\omega)}{\partial \omega}\right)\pi
    A_\sigma(\omega),
    \end{equation}
where $f(\omega)$ denotes the Fermi-Dirac distribution function,
while $A_\sigma(\omega)$ is the spin-dependent \emph{spectral
function} of the OL,
    \begin{equation}
    A_\sigma(\omega)
    =
    -\frac{1}{\pi}
    \text{Im}\,
    \langle\!\langle
     \hat{c}_\sigma^{}
    |
     \hat{c}_\sigma^\dagger
    \rangle\!\rangle_\omega^\text{r}
    .
    \end{equation}
In the equation above,
    $
    \langle\!\langle
     \hat{c}_\sigma^{}
    |
     \hat{c}_\sigma^\dagger
    \rangle\!\rangle_\omega^\text{r}
    $
represents the Fourier transformation of the retarded Green's function
    $
    \langle\!\langle
     \hat{c}_\sigma^{}
    |
     \hat{c}_\sigma^\dagger
    \rangle\!\rangle_\omega^\text{r}
    =
    -i\theta(t)
    \langle\{ \hat{c}_\sigma^{}(t), \hat{c}_\sigma^\dagger(0)\}\rangle
    $
of the orbital level.
To determine the spectral function $A_\sigma(\omega)$,
we use the Wilson's numerical renormalization group (NRG) method.~\cite{Wilson_Rev.Mod.Phys.47/1975,Bulla_Rev.Mod.Phys.80/2008}
Specifically, the recent idea of a full density
matrix,~\cite{Weichselbaum_Phys.Rev.Lett.99/2007} which allows for
reliable calculation of static and dynamic properties of the
system at arbitrary temperatures, is employed.~\cite{Legeza_DMNRGmanual,Toth_Phys.Rev.B78/2008}
For the present problem, the $U_\textrm{charge}(1)$ symmetry was exploited,
the discretization parameter $\Lambda=1.8$ was used
and we kept $N_k=2500$ states during calculations.

\section{Results and discussion}
\subsection{Parameters}

In our considerations we assume the following model parameters:
$U/W=0.4$, $\Gamma/U=0.1$, $|J|/U = 0.01125$, with $W\equiv 1$ being the energy unit,
while the spin polarization of the leads is $P=0.5$.
Without loss of generality, we assume that the molecule
is characterized by a hypothetical spin $S=2$.
In the case of $J=0$, the Kondo temperature
$T_\text{K}$ (expressed in units of energy, $k_\text{B}\equiv 1$)
of the system for nonmagnetic leads and
for $\varepsilon = -U/2$, is $T_\text{K}^0 / W \approx 0.002$,
which will be used as a reference value.
In this paper $T_\text{K}$ is extracted from the temperature
dependence of the total linear conductance
as the value of temperature $T$ at which $G(T )/G(T=0) = 1/2$.

\subsection{Ground state properties}

Generally, the Kondo effect can arise in the system when the OL is occupied
by a single electron and temperature $T$ is lower than the
Kondo temperature $T_\text{K}$.
Due to the exchange interaction $J$ between the spin of an electron in the OL
and the magnetic core effective spin, the MQD's magnetic states
decompose into two spin multiplets, characterized by the total spin number $S\pm1/2$,
whose relative position is governed by the sign of~$J$.
At sufficiently low temperatures, the transport properties of the system
can be entirely determined by its ground state.
Consequently, in order to gain a better understanding of the role
the transverse anisotropy plays in transport properties of the system,
it may be instructive to analyze first the ground state
of an \emph{isolated} MQD in two specific cases:
(1) when the core exhibits only \emph{uniaxial} magnetic anisotropy,
and
(2) when also the \emph{transverse} component is present.
Note that, although in the following discussion we address an \emph{integer} spin $S$,
analogous analysis can be carried out also for a system with a \emph{half-integer} spin $S$.

\subsubsection{No transverse anisotropy ($E=0$)}

In the absence of external magnetic field, $|\vec{B}|=0$,  the Hamiltonian~(\ref{Eq:Ham_MQD}) can be diagonalized analytically and its eigenstates enumerated with the eigenvalues $M$ of  the $z$th component $ \hat{S}_z^t$ of the total spin operator $ \hat{\vec{S}}^t= \hat{\vec{S}}+ \hat{\vec{s}}$.~\cite{Timm_Phys.Rev.B73/2006,Misiorny_Phys.Stat.Sol.B246/2009} As a result, the degenerate ground state dublets for the spin multiplet $S+1/2$ (labelled as `FM') are found to be
    \begin{equation}\label{Eq:GS_FM}
        \ketin{\pm S\pm\tfrac{1}{2}}{FM}
        =
        \phi_{\pm S\pm1/2}^{\uparrow(\downarrow),\pm S}\,
        \ketin{\!\uparrow\!(\downarrow)}{OL}\!\otimes\!\ketin{\pm S}{core},
    \end{equation}
and for the spin multiplet $S-1/2$ (labelled as `AFM')
        \begin{align}\label{Eq:GS_AFM}
        \ketin{\pm S\mp\tfrac{1}{2}}{AFM}
        &=
        \psi_{\pm S\mp 1/2}^{\downarrow(\uparrow),\pm S}\,
        \ketin{\!\downarrow\!(\uparrow)}{OL}\!\otimes\!\ketin{\pm S}{core}
                \nonumber\\
        &+
        \psi_{\pm S\mp1/2}^{\uparrow(\downarrow),\pm S\mp1}\,
        \ketin{\!\uparrow\!(\downarrow)}{OL}\!\otimes\!\ketin{\pm S\mp1}{core},
        \end{align}
where $\ketinline{\bullet}{OL\,(core)}$ denotes the spin state of the OL (magnetic core), and
    $
    \phi_{M}^{\sigma,m} (\psi_{M}^{\sigma,m})
    \equiv
    \Big[
    \brainline{\sigma}{OL}
    \otimes
    \brainline{m}{core}
    \Big]
    \ketinline{M}{FM\,(AFM)}
    $
represents the overlap of state
$\ketinline{\sigma}{OL}\otimes\ketinline{m}{core}$
with the eigenstate
$\ketinline{M}{FM\,(AFM)}$.
In equations above $\sigma$ stands for the spin index of an electron in the OL, while $m$ is the eigenvalue of the internal spin operator $ \hat{S}_z$ so that $|m|=0,\ldots,S$. It goes without saying that in Eq.~(\ref{Eq:GS_FM}) there must be $\phi_{\pm S\pm1/2}^{\uparrow(\downarrow),\pm S}=1$, whereas the explicit expressions for $\psi_{M}^{\sigma,m}$ can be found, for instance, in Ref.~[\onlinecite{Misiorny_Phys.Stat.Sol.B246/2009}].

\subsubsection{With transverse anisotropy ($E\neq0$)}

For a finite component of transverse magnetic anisotropy the relatively simple picture for the MQD's ground state developed above is no longer applicable. At present, each of the $2S+1$ core-spin states
$\ketinline{\chi}{core}$,
$ \hat{\mathcal{H}}_\text{S}\ketinline{\chi}{core}\!=\!E_\chi\ketinline{\chi}{core}$,
can be a linear combination of the $S_z$-eigenstates $\ketinline{m}{core}$.
In order to keep the notation transparent, let's introduce an auxiliary subscript~$\nu$, i.e., $\ketinline{\chi}{core}\!\rightarrow\!\ketinline{\chi_\nu}{core}$,
and assume that $|\nu|=0,\ldots,S$.
Then, if $E\neq0$ and $|\vec{B}|=0$, one can use the expansion
    $
    \ketin{\chi_\nu}{core}
    \!=\!
    \sum_{\xi}\leftidx{_{\raisebox{-1pt}{\scalebox{0.6}{\textrm{core}}}}\!}{\langle}\nu+2\xi\ketinline{\chi_\nu}{core}
    \,\ketinline{\nu+2\xi}{core},
    $
with the summation running over integer $\xi$ satisfying $|\nu+2\xi|\leqslant S$.
As a result, one observes that each of the states
$\ketinline{\chi_\nu}{core}$
is formed from states belonging exclusively to one of
two uncoupled sets:
$\big\{l=0,1,\ldots,S\!\!: \ketinline{S-2l}{core}\big\}$
and
$\big\{l=0,1,\ldots,S-1\!\!: \ketinline{S-1-2l}{core}\big\}$,~\cite{Romeike_Phys.Rev.Lett.96/2006,Romeike_Phys.Rev.Lett.96/2006a} i.e., each grouping states
$\ketinline{m}{core}$ with the same parity with respect to $m$. Furthermore, by analyzing how the Hamiltonian $\mathcal{H}_\textrm{MQD}$ acts on the states $\ketinline{\sigma}{OL}\!\otimes\!\ketinline{\chi_\nu}{core}$,
one can deduce the corresponding form of the MQD's ground state doublets:
        \begin{align}\label{Eq:GS_FM_E}
        \ketin{\chi_{\pm S\pm1/2}}{FM}
        =
        \sum_m
        \Big\{
        &
        \phi_{\chi_{\pm S\pm1/2}}^{\uparrow,m}\,
        \ketin{\!\uparrow}{OL}\!\otimes\!\ketin{m}{core}
                       \nonumber \\[-5pt]
        +\ &
        \phi_{\chi_{\pm S\pm1/2}}^{\downarrow,m}\,
        \ketin{\!\downarrow}{OL}\!\otimes\!\ketin{m}{core}
        \Big\},
        \end{align}
        \begin{align}\label{Eq:GS_AFM_E}
        \ketin{\chi_{\pm S\mp1/2}}{AFM}=
        \sum_m
        \Big\{
        &
        \psi_{\chi_{\pm S\mp1/2}}^{\downarrow,m}\,
        \ketin{\!\downarrow}{OL}\!\otimes\!\ketin{m}{core}
            \nonumber\\[-5pt]
        +\ &
        \psi_{\chi_{\pm S\mp1/2}}^{\uparrow,m}\,
        \ketin{\!\uparrow}{OL}\!\otimes\!\ketin{m}{core}
        \Big\}
        .
        \end{align}
Note that now the subscript $\nu$ in $\ketinline{\chi_\nu}{FM\, (AFM)}$
has a clear meaning, namely it stands for the $M$ component of highest weight in the state
$\ketinline{\chi_\nu}{FM\, (AFM)}$, so that  $\ketinline{\chi_M}{FM\, (AFM)}\equiv\ketinline{M}{FM\, (AFM)}$
 for $E\rightarrow0$.
In the present case, there are no general explicit formulae for the
coefficients $\phi_{\chi_{M}}^{\sigma,m}$ and $\psi_{\chi_{M}}^{\sigma,m}$
for an arbitrary spin number $S$, so that these have to be obtained numerically.
In addition, it is worth emphasizing that since $S$ is integer, so that $S^t$ is half-integer when the OL is occupied by a single electron, the both ground states
are still twofold degenerate (Kramers' doublets).
Consequently, one should expect the Kondo effect to occur
at sufficiently low temperatures.

\subsection{Spectral functions}

\begin{figure}
\centering
\includegraphics[width=0.95\columnwidth]{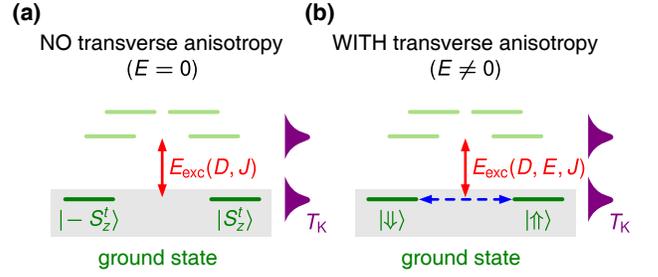}
\caption{(color online)
Symbolic illustration of the effect of transverse magnetic anisotropy on the MQD's ground state.
At low temperatures for $D,|J|>T_\text{K}$ and $E=0$ (a), the transport
properties of the system are fully determined by its ground state
doublet $|\pm S_z^t\rangle$, which is then energetically well separated from the first excited doublet,
with the excitation energy $E_\text{exc}$ depending both on the magnetic core's anisotropy
and its exchange interaction with the electron spin in the OL.
Note that transitions between the states of the doublet are not permitted.
For $E\neq0$ (b), on the other hand, such transitions can occur
(indicated by the double-arrow dashed line) and the system effectively
behaves as a one-a-half pseudospin.
}
\label{Fig:2}
\end{figure}

It has already been shown that if the magnetic core of MQD exhibits
the \emph{uniaxial} component of magnetic anisotropy,
it can lead to the suppression of the Kondo effect.~\cite{Misiorny_Phys.Rev.Lett.106/2011,Misiorny_Phys.Rev.B84/2011}
Importantly, the occurrence of the Kondo effect is conditioned by
the competition between the energy scales set by the exchange coupling
$J$ and $T_\text{K}^0$ (the Kondo temperature in the case of $J=0$).
For  $|J|<T_\text{K}^0$, the system minimizes its energy by forming
the many-body Kondo state as a result of strong hybridization
between an electron in the OL and free electrons in electrodes,
whereas for $|J|>T_\text{K}^0$ the electron's spin couples
\emph{via} exchange interaction with the core spin.
In the latter situation, the type of exchange interaction plays a crucial role.
In the case of vanishing anisotropy, for ferromagnetic exchange coupling $J$,
one always observes the Kondo effect at sufficiently low temperatures,
\cite{Kusunose_J.Phys.Soc.Jpn.66/1997}
while for antiferromagnetic $J$, the system exhibits a two-stage Kondo effect
as a function of temperature.
\cite{Vojta_PhysRevB.65.140405}
On the other hand, for finite magnetic anisotropy, the above effects
can be suppressed once $|D| \gtrsim T_{\rm K}$.
\cite{Misiorny_Phys.Rev.B86/2012_UK,Misiorny_Phys.Rev.B86/2012}
Here, we are in particular interested in the effects resulting from transverse magnetic anisotropy,
therefore in the following we assume $|J|>T_\text{K}^0$.
We also set the uniaxial anisotropy to be $D/T^0_{\rm K} = 0.75$, unless stated otherwise.

Besides the energy scales discussed above, in the case of ferromagnetic leads
the occurrence of the Kondo effect is conditioned by the magnitude of
the ferromagnetic-contact-induced dipolar exchange field $\exch$.
\cite{Martinek_Phys.Rev.Lett.91/2003_127203,Martinek_Phys.Rev.Lett.91/2003_247202,
Choi_Phys.Rev.Lett.92/2004}
The Kondo resonance is then suppressed when $\exch\gtrsim T_{\rm  K}$.
\cite{Pasupathy_Science306/2004,Hauptmann_NaturePhys.4/2008,Gaass_Phys.Rev.Lett.107/2011}
$\exch$ results directly from spin-dependence of the couplings between the OL
and the leads. For symmetric systems, in the antiparallel configurations,
the resultant coupling does not depend on spin and
the exchange field develops only in the parallel magnetic configuration,
with sign and magnitude controllable by the gate voltage,
$\exch \propto \log |\varepsilon / (\varepsilon+U)|$.

\begin{figure}[t!]
\begin{center}
\includegraphics[width=0.95\columnwidth]{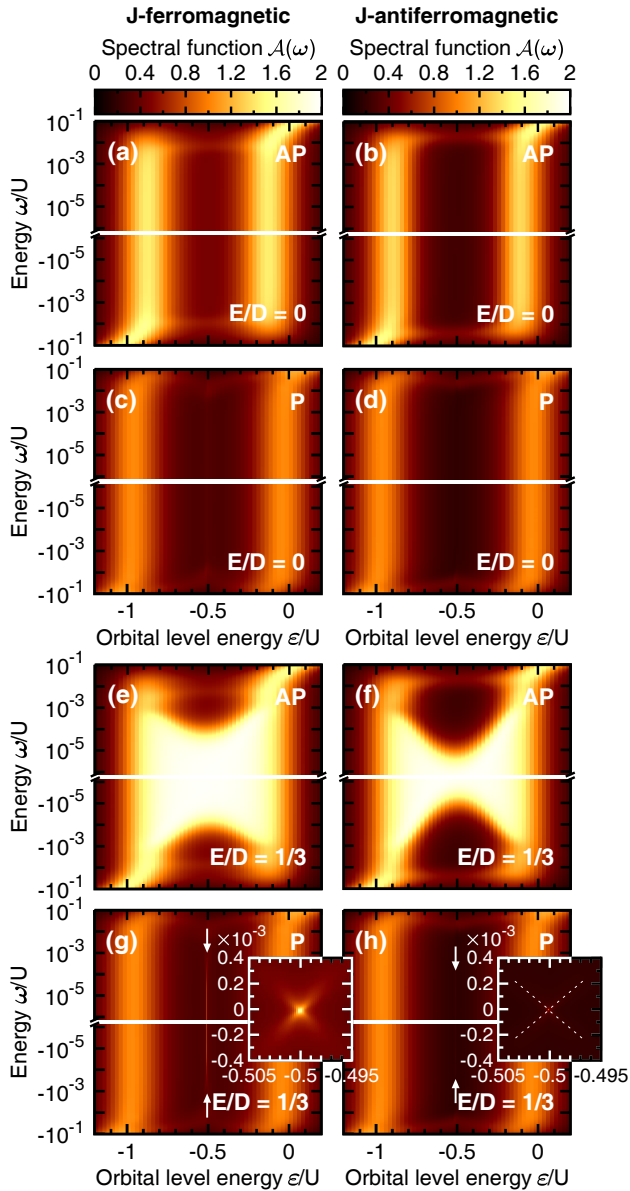}
\end{center}
\caption{
Dependence of total \emph{normalized} orbital level (OL) spectral function,
$\mathcal{A}(\omega)=\pi\sum_\sigma \Gamma_\sigma A_\sigma(\omega)$,
on energy $\omega$ and the OL position $\varepsilon$
for ferromagnetic (left column) and antiferromagnetic (right column)
exchange coupling in the case of (a)-(d) vanishingly small,
and (e)-(h) large value of transverse magnetic anisotropy $E$.
The first and third rows correspond to the antiparallel magnetic configuration of electrodes,
while the second and fourth rows show the case of parallel magnetic configuration.
Insets in the bottom panels are magnifications of spectral functions
in the parameter space around the particle-hole symmetry point ($\varepsilon=-U/2$),
with the color scale kept the same as in other panels.
Note that since the $\omega$- and $\varepsilon$-scales in both
insets are assumed the same, the restoration of the
Kondo resonance in inset to (h) is hardly visible, though it occurs.
Moreover, dashed lines as a guide for eyes are added in inset (h)
to highlight the resonance splitting due to the dipolar exchange field.
Key parameters are given  in the main text with
$D/U=3.75\times10^{-3}$ ($D/T_\textrm{K}^0=0.75$)
and $E$ is specified in the bottom-right corner of each panel.
}
\vspace*{-30pt}
\label{Fig:3}
\end{figure}

Figure \ref{Fig:2}  depicts symbolically the effect of transverse anisotropy
on the ground state of the system.
At low temperatures and for $D,|J|>T_\text{K}$ and $E=0$, Fig.~\ref{Fig:2}(a),
the ground state of the system is a doublet $|\pm S_z^t\rangle$,
which is energetically well separated from the other excited states,
so that the transitions between states of the doublet are not permitted.
In the case of finite $E$, Fig.~\ref{Fig:2}(b),  such transitions can occur
and the system effectively behaves as an $S=1/2$ pseudospin.
As a result, in the presence of transverse anisotropy, the system
can exhibit the Kondo effect.
This behavior can be observed in the dependence of the normalized spectral functions
$\mathcal{A}(\omega)=\pi\sum_\sigma \Gamma_\sigma A_\sigma(\omega)$
on energy~$\omega$ (note the logarithmic scale)
and the OL position~$\varepsilon$ in the case of both parallel
and antiparallel magnetic configurations, shown in Fig. \ref{Fig:3}.

\begin{figure}[t]
\begin{center}
\includegraphics[width=0.8\columnwidth]{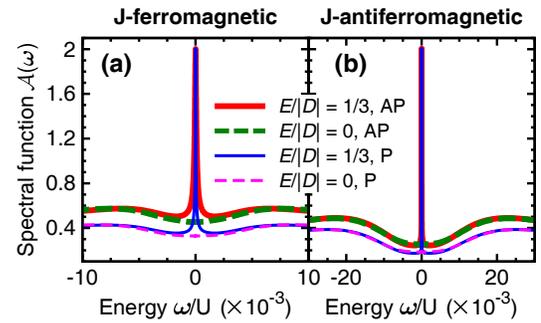}
\end{center}
\caption{
The normalized spectral functions at the particle hole symmetry point,
$\varepsilon = -U/2$, in both magnetic configurations
and for $E/D = 0$ (dashed lines) and $E/D=1/3$ (solid lines). The left panel corresponds
to the case of \emph{ferromagnetic} ($J>0$) exchange coupling,
while the right panel  corresponds to the case of
\emph{antiferromagnetic} ($J<0$) exchange interaction.
The other parameters are the same as in Fig. \ref{Fig:3}.
}
\label{Fig:4}
\end{figure}

For finite $D$ and $E=0$, the Kondo effect becomes generally suppressed,
irrespective of the magnetic configuration and the sign of the exchange coupling,
see Figs.~\ref{Fig:3}(a)-(d).
The origin of this suppression can be qualitatively understood as follows.
Let us for simplicity consider the large-spin ground state doublet
of a bare MQD, given by Eqs.~(\ref{Eq:GS_FM})-(\ref{Eq:GS_AFM}).
One then finds that electron cotunneling processes
that can result in reversing the spin of an electron in the OL
do not permit direct transitions within the doublet
(effectively corresponding to spin-exchange processes for the MQD's total spin),
regardless of the sign of $J$, meaning that the many-body Kondo
state cannot be formed, Fig.~\ref{Fig:2}(a).
Furthermore, the suppression is more pronounced for the AFM $J$-coupling.
This stems from the differences in energies and forms of the ground
state doublets for FM and AFM cases. In particular, since the ground state
energy for $\ketinline{\pm S\mp1/2}{AFM}$ is lower than
that for $\ketinline{\pm S\mp1/2}{FM}$, with the energies of virtual states
for singly and doubly occupied OL being independent of~$J$, and,
unlike the state $\ketinline{\pm S\mp1/2}{FM}$,
the state $\ketinline{\pm S\mp1/2}{AFM}$ involves a superposition of both
OL spin states `up' and `down', cf. Eqs.~(\ref{Eq:GS_FM})-(\ref{Eq:GS_AFM}),
these translate into less efficient cotunneling processes driving linear transport for the AFM $J$-coupling.
Moreover, the suppression of the conductance in the local moment regime,
$-U<\varepsilon< 0$, is more pronounced in the parallel configuration as compared
to the antiparallel configuration, which is related with the presence of the exchange field.

Interestingly enough, when the transverse component of magnetic anisotropy is additionally included,
the situation changes dramatically, as this component [see the second term of Eq.~(\ref{Eq:Ham_S})]
makes  mixing of the core spin states possible, Fig.~\ref{Fig:2}(b).
In consequence, each of the states belonging to the FM/AFM ground state
doublet becomes now a superposition of all available OL electron spin states
$\ketinline{\sigma}{OL}$ and core spin states $\ketinline{m}{core}$,
see Eqs.~(\ref{Eq:GS_FM_E})-(\ref{Eq:GS_AFM_E}).
Since
$\phi_{\chi_{\pm S\pm1/2}}^{\uparrow(\downarrow),m}\phi_{\chi_{\mp S\mp1/2}}^{\downarrow(\uparrow),m}\neq0$
($\psi_{\chi_{\pm S\mp1/2}}^{\uparrow(\downarrow),m}\psi_{\chi_{\mp S\pm1/2}}^{\downarrow(\uparrow),m}\neq0$),
the effective spin-exchange processes for the MQD's total spin owing to the OL electron cotunneling are allowed,
and the ground state doublet can effectually be viewed as the pseudospin-1/2 system.
This, in turn, manifests as a revival of the Kondo effect
in the case of finite $E$, as one can see in Figs.~\ref{Fig:3}(e)-(h).
It is also importnant to note a large quantitative difference between the parallel
and antiparallel configurations in the case of finite transverse anisotropy.
While in the antiparallel configuration the Kondo resonance is restored
in the whole Coulomb blockade regime with a single electron in OL,
in the parallel configuration, the Kondo resonance occurs only
at the particle-hole symmetry point, that is at the point where the dipolar exchange field cancels.
Moreover, the Kondo temperature in the parallel configuration is
much smaller than that in the antiparallel configuration,
see the insets in Figs.~\ref{Fig:3}(g)-(h), which zoom into the
low-energy regime around the particle-symmetry point, $\varepsilon = -U/2$.
This is because the effective exchange coupling between the spin
in the MQD and the spins of conduction electrons is lowered by a
lead-spin-polarization dependent factor smaller than unity.
\cite{Martinek_Phys.Rev.Lett.91/2003_127203}

The difference between the cases of vanishing and finite transverse anisotropy
constant can be explicitly seen in Fig.~\ref{Fig:4}, which
shows the normalized spectral functions calculated for $\varepsilon = -U/2$
for both magnetic configurations in the case of ferromagnetic and antiferromagnetic~$J$
(i.e., the relevant cross-sections from Fig.~\ref{Fig:3}).
Clearly, irrespective of the sign of the exchange coupling $J$, finite transverse
anisotropy restores the Kondo effect. The restoration can be observed
in both magnetic configurations with the Kondo temperature much smaller
in the parallel configuration compared to the antiparallel one.

\subsection{The linear conductance and TMR}

The subtle interplay between all the energy scales is also visible in
the behavior of the linear conductance $G$.
In addition, to describe the change of system transport properties
when the magnetic configuration is varied between parallel and antiparallel,
we study the behavior of the \emph{tunnel magnetoresistance}, which is defined as
\cite{Julliere_Phys.Lett.A54/1975}
    \begin{equation}
    \text{TMR}
    =
    \frac{
    G_\text{P}-G_\text{AP}
    }{
    G_\text{AP}
    }
    ,
    \end{equation}
where $G_{\rm P}$ ($G_{\rm AP}$) represents  the linear conductance in the parallel (antiparallel) configuration.

\subsubsection{Orbital level dependence}

\begin{figure}[t]
\includegraphics[width=0.99\columnwidth]{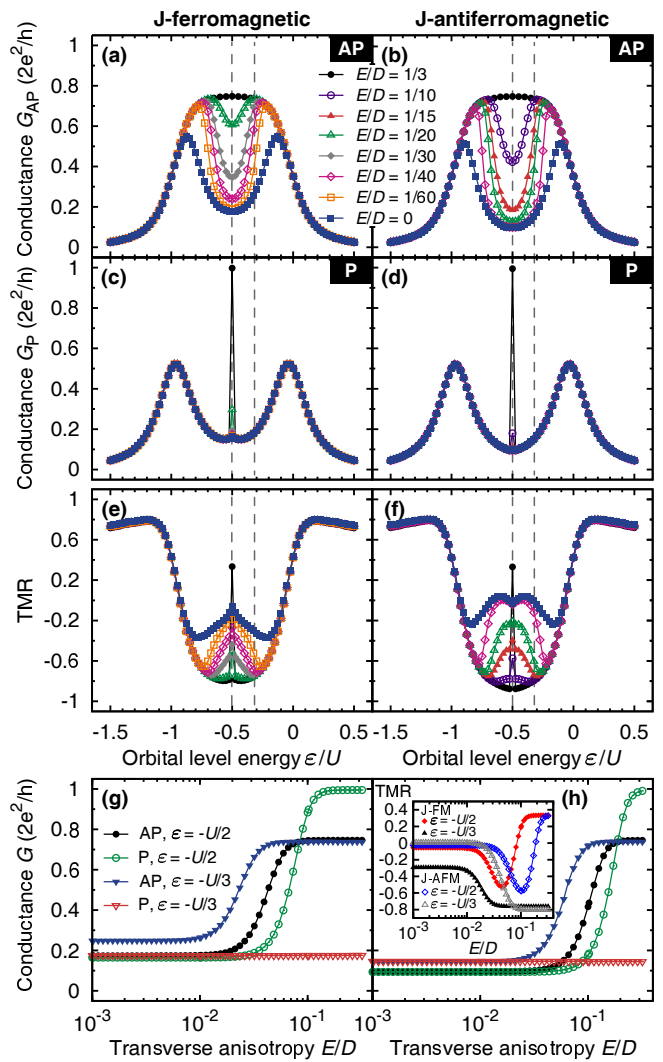}
\caption{(color online)
The linear conductance $G$ for the (a)-(b)
\emph{antiparallel} (AP) and (c)-(d) \emph{parallel} (P)
magnetic configuration, and the corresponding TMR (e)-(f),
shown as a function of the OL energy $\varepsilon$ for different values of
the transverse anisotropy constant $E$.
(g)-(h)
The dependence of $G_{\rm P}$ and $G_{\rm AP}$ on the transverse
magnetic anisotropy (scaled with respect to the uniaxial anisotropy)
for $\varepsilon = -U/2$ and $\varepsilon=-U/3$, see vertical dashed lines in (a)-(f).
Corresponding TMR is shown in inset to (h).
\emph{Left/right} column represents the case of \emph{ferromagnetic}
($J>0$)/\emph{antiferromagnetic} ($J<0$) exchange interaction.
The parameters are as in Fig.~\ref{Fig:3} with
$T/W=10^{-8}$ ($T/T_\textrm{K}^0=5\times10^{-6}$).
}
\label{Fig:5}
\end{figure}

The orbital level dependence of $G_{\rm P}$, $G_{\rm AP}$ and TMR
is shown in Fig.~\ref{Fig:5} for both ferromagnetic and antiferromagnetic
exchange coupling $J$ and for different values of the transverse anisotropy constant $E$.
In the case when the orbital level is empty ($\varepsilon>0$) or doubly occupied
($\varepsilon<-U$) the coupling to the core spin does not play any role
and the conductance does not depend on $E$.
This is contrary to the case when the OL is singly occupied
($-U<\varepsilon<0$), see Fig.~\ref{Fig:5}.
In the antiparallel configuration of electrodes' magnetic moments,
Fig.~\ref{Fig:5}(a)-(b), the conductance is then suppressed
for $E=0$ and increases with increasing $E$ to its maximum value
of $(2e^2/h)(1-P^2)$.

On the other hand, in the parallel magnetic configuration,
Fig.~\ref{Fig:5}(c)-(d), due to the presence of the exchange field,
the ground state doublet is split, so that the influence of magnetic anisotropy
is limited only to the particle-hole symmetry point ($\varepsilon=-U/2$)
where the field disappears, and the conductance can reach
the limit value of the conductance quantum $2e^2/h$.
Furthermore, with increasing the transverse magnetic anisotropy
constant towards its maximal value, i.e., $E/D\rightarrow 1/3$,
one observes that the differences between the cases of FM and AFM $J$-coupling
are almost indistinguishable. In fact, in this limit transport
signatures of the MQD start resembling these typical
to a single-level quantum dot, \cite{Weymann_Phys.Rev.B83/2011}
that is corresponding to a MQD in the limit of vanishingly small
exchange interaction $J\rightarrow0$.

The difference between the linear conductance in
the two magnetic configurations of the system is reflected in the TMR,
which is shown in Fig.~\ref{Fig:5}(e)-(f).
For large transverse anisotropy, $E/D=1/3$, the TMR for $\varepsilon = -U/2$
is given by $P^2 / (1-P^2)$, while for
$-U < \varepsilon <0$ and $\varepsilon \neq -U/2$, $G_{\rm P}$
is suppressed by the exchange field and ${\rm TMR} \to -1$.
However, for smaller transverse anisotropy, $|\rm TMR|$ is decreased
for both positive and negative exchange interaction $J$. This is because
$G_{\rm AP}$ drops with decreasing $E$, while $G_{\rm P}$ does not depend on
$E$ for $\varepsilon\neq -U/2$, i.e., for such level position where the exchange field
is present.

The explicit dependence of linear conductance and TMR on $E$
is shown in Fig.~\ref{Fig:5}(g)-(h).
It can be seen that the precise value of transverse magnetic anisotropy for which
the conductance reaches its maximum value depends on parameters of the model.
Quite generally, the Kondo effect is already well restored for $E/D\gtrsim 1/5$.
The resulting dependence of TMR on $E$ is presented in the inset
to Fig. \ref{Fig:5}(h). For $\varepsilon = -U/2$, the TMR exhibits a minimum
for such $E$ where the conductance starts increasing and then increases to
the value of $P^2 / (1-P^2)$. On the other hand, for $\varepsilon =-U/3$,
the TMR decreases with increasing $E$ reaching large negative value
with ${\rm TMR}\approx -1$.
This basically means that the MQD conducts better in the antiparallel magnetic configuration,
which stems from the presence of the dipolar exchange field
in the parallel configuration in the case under consideration.

\subsubsection{Temperature dependence}

\begin{figure}[t!!!]
\includegraphics[width=0.99\columnwidth]{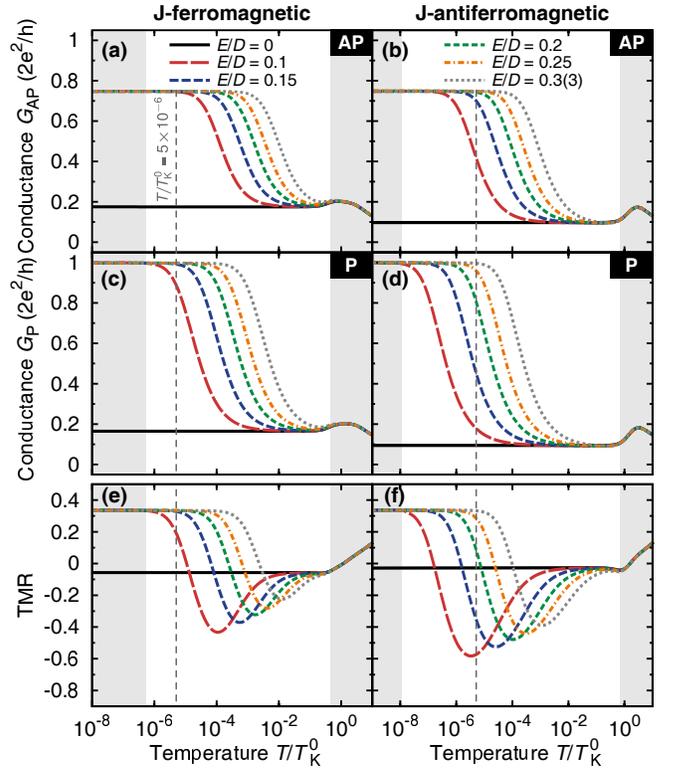}
\caption{(color online)
The temperature dependence of the linear conductance $G$
in the antiparallel (a)-(b) and parallel (c)-(d) configuration
and the resulting TMR (e)-(f)
for several values of the transverse anisotropy
parameter $E$ and for $\varepsilon=-U/2$.
Solid lines represent the case when the
transverse magnetic anisotropy is absent ($E=0$).
Vertical dashed lines represent the temperature used in~Fig.~\ref{Fig:5},
$T/T_\textrm{K}^0=5\times10^{-6}$.
Remaining parameters as in Fig.~\ref{Fig:3}.
}
\label{Fig:6}
\end{figure}

\begin{figure}[t!!!]
\includegraphics[width=0.99\columnwidth]{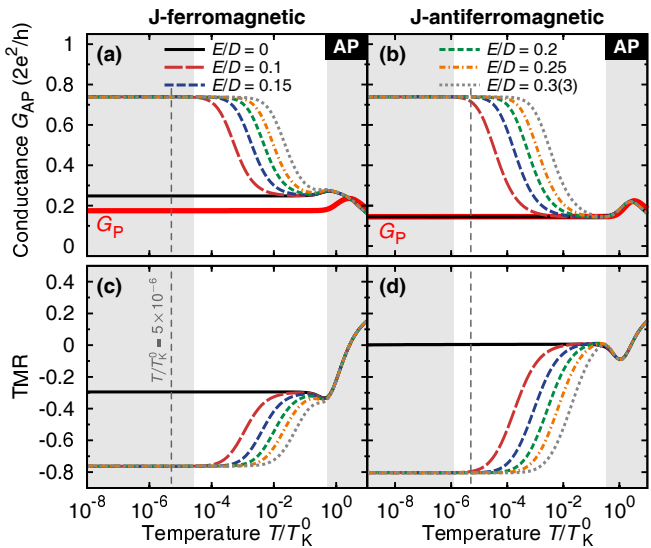}
\caption{(color online)
Analogous to  Fig.~\ref{Fig:6}, but  for $\varepsilon=-U/3$.
Note that due to the presence of the effective exchange
field for the parallel magnetic configuration
the restoration of the Kondo effect does not take place,
so that only the results for the antiparallel magnetic configuration (a)-(b) are of main interest.
Nevertheless, for the sake of completeness,
bold solid lines representing conductance $G_\text{P}$
in the parallel configuration are also plotted.
In order to enable comparison with the case of $\varepsilon=-U/2$,
the range of scales for all axes is assumed the same as in Fig.~\ref{Fig:6}.
Remaining parameters as in Fig.~\ref{Fig:3}.
}
\label{Fig:7}
\end{figure}

Although in Fig.~\ref{Fig:5} the complete restoration
of the Kondo effect occurs only for $E/D \gtrsim 1/5$, see Fig.~\ref{Fig:5},
one should bear in mind that these results have been obtained for a specific, finite temperature.
In fact, it turns out that the effect can be reinstated for any $E \neq 0$
with the Kondo temperature depending now on $E$.
Figures~\ref{Fig:6} and~\ref{Fig:7} illustrate the temperature dependence
of the linear conductance for indicated values of the transverse anisotropy constant $E$
and two distinctive values of the OL energy: $\varepsilon= -U/2$
(Fig.~\ref{Fig:6}) and $\varepsilon= -U/3$ (Fig.~\ref{Fig:7}).
The essential difference between these two cases stems from
the absence (presence) of the effective exchange field for  $\varepsilon= -U/2$
($\varepsilon= -U/3$) in the parallel configuration,
which is reflected in the behavior of TMR.
In particular, this effective field leads to the splitting of the ground state doublet,
precluding in consequence the formation of the Kondo effect for $\varepsilon= -U/3$.

\begin{figure}[t]
\includegraphics[width=0.6\columnwidth]{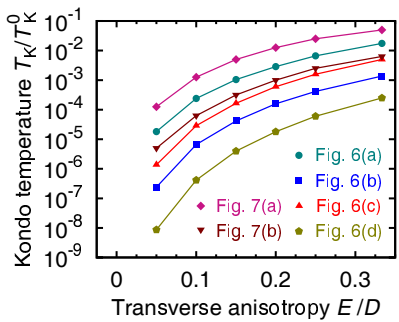}
\caption{(color online) Kondo temperature as a function
of the transverse magnetic anisotropy constant $E$,
estimated from the temperature dependence of $G$
for different magnetic configurations and different types of the $J$-coupling
presented in Figs.~\ref{Fig:6} and~\ref{Fig:7}.
}
\label{Fig:8}
\end{figure}

First of all, one can notice that Kondo temperatures observed
in the situation when the Kondo effect originates from
the transverse magnetic anisotropy are generally lower
than the reference Kondo temperature $T_\text{K}^0$
of a single-level quantum dot (that is for $J=0$), see Fig.~\ref{Fig:8}.
Furthermore, at the particle-hole symmetry point ($\varepsilon=-U/2$)
these temperatures depend also on the magnetic configuration of electrodes,
being lower for the parallel configuration [cf. (a)-(b) with (c)-(d) in Fig.~\ref{Fig:6}].
Such disparity, in turn, reveals clearly as a
nonmonotonic dependence of TMR on temperature $T$.

In the limit of \emph{larg}e $T$, $D\gg T\gg T_\text{K}$,
the system remains in the low-conducting state with the
value of conductance being insensitive to the presence
of the transverse magnetic anisotropy. On the contrary,
its presence becomes clearly visible in the limit of \emph{low} $T$,
$ T\ll T_\text{K}$, where for $E \neq 0$ the system enters
the high-conducting state due to the Kondo effect,
provided that the ground state doublet is not affected
by the effective exchange field. In the present situation,
in particular, it implies that no Kondo effect, and accordingly no high-conducting state,
should generally be expected in the parallel configuration
except for $\varepsilon=-U/2$, cf. Fig.~\ref{Fig:6}(c)-(d)
and solid lines marked as $G_\text{P}$ in Fig.~\ref{Fig:7}(a)-(b).
As a result, one observes that for $E\neq0$ both in the limit
of \emph{large} and \emph{low} temperature,
marked in Figs.~\ref{Fig:6}-\ref{Fig:7} as shaded areas,
TMR takes constant asymptotic values which are independent of the actual value of $E$.

On the other hand, the presence of transverse magnetic anisotropy
manifests in TMR for the intermediate range of temperatures
with respect to the limiting cases discussed above.
For the particle-hole symmetry point $\varepsilon=-U/2$, Fig.~\ref{Fig:6}(e)-(f),
one observes then a global minimum in TMR to develop.
This reflects the fact that for a given value of $E$ the
Kondo temperature $T_\text{K}$ differs for the antiparallel
and parallel magnetic configuration of electrodes,
with $T_\text{K}$ being generally lower in the latter case.
The reason for this difference was explained in the previous section.
A qualitatively different behavior of TMR is seen for $\varepsilon=-U/3$,
Fig.~\ref{Fig:7}(c)-(d), where the temperature dependence
of TMR is dominated by the monotonic increase in the
temperature range under discussion with only negative values observed.
The main reason for this is the presence of the effective exchange
field in the parallel magnetic configuration,
which, by splitting the ground state doublet, prevents the Kondo effect from taking place,
Fig.~\ref{Fig:7}(a)-(b). In consequence, neither the conductance
$G_\text{P}$ reaches the unitary limit at low temperatures
nor it depends on $E$.

Finally, in Fig.~\ref{Fig:8} values of the Kondo temperature $T_\text{K}$ derived from the $T$-dependencies of conductance shown in Figs.~\ref{Fig:6}-\ref{Fig:7} are presented as a function of the transverse magnetic anisotropy constant $E$. One can see that not only does $T_\text{K}$ generally decrease with lowering $E$, but also its value substantially depends on the type of the $J$-coupling. Specifically, one obtains larger values of $T_\text{K}$ in the FM case ($J>0$). Moreover, at the particle-hole symmetry point $\varepsilon=-U/2$, Fig.~\ref{Fig:6}, for a given $E$ larger Kondo temperature is observed for the antiparallel magnetic configuration of electrodes, but, on the other hand,  it is lower that its value in the corresponding case for $\varepsilon=-U/3$,  Fig.~\ref{Fig:7}.

\subsection{Uniaxial vs. transverse magnetic anisotropy}

\begin{figure}[t]
\includegraphics[width=0.99\columnwidth]{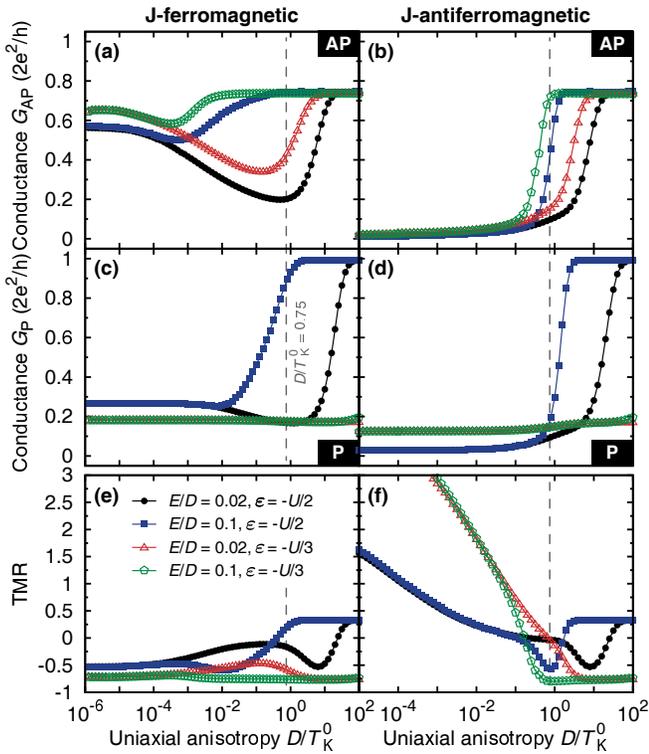}
\caption{(color online)
The influence of the interplay of \emph{uniaxial} ($D$) and \emph{transverse} ($E$)
magnetic anisotropy on transport properties of the system.
The linear conductance $G$ (a)-(d) and TMR (e)-(f) are
shown as a function of $D$ for two chosen values of the ratio $E/D$,
with \emph{full} (\emph{open}) points corresponding to $\varepsilon=-U/2$ ($\varepsilon=-U/3$).
Vertical dashed lines indicate the value of $D$ used so far.
The parameters are as in Fig.~\ref{Fig:3} with
$T/T_\text{K}^0=5\times10^{-6}$.
}
\label{Fig:9}
\end{figure}

Up to this point, the discussion has been based
on the assumption that the value of the uniaxial anisotropy parameter $D$
takes one specific value of $D/T_\text{K}^0=0.75$,
which is larger than Kondo temperature $T_\text{K}$
estimated in the presence of transverse magnetic anisotropy, see Fig.~\ref{Fig:8}.
To make the discussion complete, we relax this assumption
and analyze how the value of $D$ influences
the transport properties of MQD in the Kondo regime when $E\neq0$.
For this purpose, in Fig.~\ref{Fig:9} we plot the dependence of
the linear conductance and TMR on $D$ for two selected values of $E/D$,
both at the particle-hole symmetry point $\varepsilon=-U/2$
(full points in Fig.~\ref{Fig:9}) and away from this point
at $\varepsilon=-U/3$ (open points in Fig.~\ref{Fig:9}).

To begin with, it can be easily seen that in
the limit of strong \emph{uniaxial} anisotropy,
$D/T_\text{K}^0> 1$, the MQD exhibits transport properties
typical to the \emph{low} temperature limit discussed in the previous section.
In particular, as long as the effective exchange field is absent,
conductance achieves the unitary limit of $2e^2/h$ in the parallel configuration
and $2(1-P^2)e^2/h$ in the antiparallel configuration.
Physically, such a limit corresponds to the situation
when the ground state pseudospin doublet is very well
separated from other excited states of the spin multiplet.
Once the value of $D$ gets decreased, it is also followed
by the reduction of the linear conductance.
Interestingly enough, further diminishing of $D$
(recall that $E$ becomes also reduced, as we keep $E/D$ constant)
towards the limit of $D/T_\text{K}^0\ll 1$ results in a strikingly different
behavior of the system, which now depends on the type
of the $J$-coupling and the magnetic configuration of electrodes.

In the \emph{antiparallel} configuration and the FM ($J>0$) case,
Fig.~\ref{Fig:9}(a), the revival of the Kondo effect eventually occurs,
whereas in the case of AFM coupling ($J<0$), Fig.~\ref{Fig:9}(b),
the transport becomes almost completely suppressed.
In order to understand this effect, note first that for $D/T\ll 1$
the MQD effectively becomes \emph{spin-isotropic}. This is because thermal excitations
between neighboring spin states, allowing for overcoming the energy barrier
for spin reversal, enable indirect transitions between the ground doublet states.
Then, in the FM case, similarly as for a system of two exchange-coupled
spin-1/2 impurities,~\cite{Kusunose_J.Phys.Soc.Jpn.66/1997,Izumida_J.Phys.Soc.Jpn.67/1998}
the Kondo effect develops even though the $J$-coupling
far surpasses the hybridization $\Gamma$,
i.e., instead of stabilizing the high-spin state, the screening of the OL's spin
is preferred.~\cite{Misiorny_Phys.Rev.B86/2012}

\begin{figure}[t]
\includegraphics[width=0.99\columnwidth]{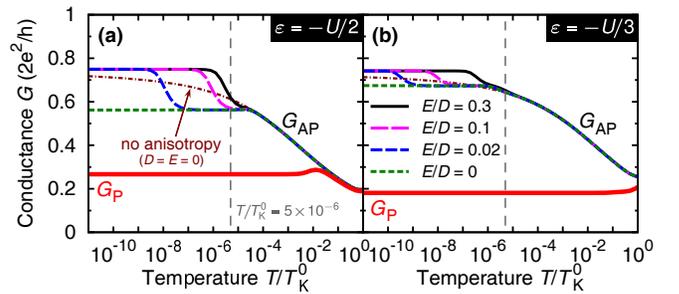}
\caption{(color online)
The temperature dependence of the linear conductance
for both magnetic configurations
in the case of $\varepsilon = -U/2$ (a) and $\varepsilon = -U/3$ (b)
for different values of the transverse anisotropy constant $E$
and $D/T_\text{K}^0=10^{-6}$.
Note that only the FM type of the $J$-coupling is considered here,
and the used value of $D/T_\text{K}^0$ corresponds
to the lowest value of $D/T_\text{K}^0$ shown in~Fig.~\ref{Fig:9}.
Moreover, vertical dashed lines indicate
the temperature for which Fig.~\ref{Fig:9} was  calculated.
All other parameters as in Fig.~\ref{Fig:3}.
}
\label{Fig:10}
\end{figure}

In order to show that the Kondo effect should indeed
be observed in Fig.~\ref{Fig:9}(a) for small values of $D$,
in Fig.~\ref{Fig:10} we present the temperature dependence
of the linear conductance $G_\text{AP}$ for $D/T_\text{K}^0=10^{-6}$.
It can be seen that, unlike for the spin-isotropic case ($D=E=0$, dashed-dotted thin lines),
the appearance of the Kondo effect when lowering temperature takes place stepwise.
To understand this behavior one should realize that
finite magnetic anisotropy generally suppresses the Kondo resonance.
For large ferromagnetic $J$-coupling, $J>T_{\rm K}^0$, as considered here,
the actual Kondo temperature is much lower than $T_{\rm K}^0$,
\cite{Misiorny_Phys.Rev.B86/2012}
such that even tiny values of $D$ can affect the low temperature
behavior of the conductance. This can be seen
in Fig.~\ref{Fig:10} in the case of vanishing transverse anisotropy,
when the conductance is smaller than its maximum value
of $G_{\rm AP} = (1-P^2)2e^2/h$.
Then, finite transverse component of magnetic anisotropy
can indeed play an important role, giving rise to full
restoration of the Kondo effect, which occurs as a step
in the dependence of $G_{\rm AP}$ on temperature.

For the AFM $J$-coupling, on the other hand, the suppression of the
conductance is expected to arise due to the inability
of screening the MQD's spin by conduction electrons.
In particular, since we assume $|J|>T_\text{K}^0$,
at low temperatures ($T\ll T_\text{K}^0$) the MQD can
be effectively viewed as a single composite spin
of value $S^t=S-1/2$ that couples ferromagnetically to the conduction band,
\cite{Zitko_J.Phys.:Condens.Matter22/2010} so that its screening becomes impossible.

For the \emph{parallel} magnetic configuration, see Fig.~\ref{Fig:9}(c)-(d),
an analogous dependence of conductance as analyzed
above can be observed for the OL energy $\varepsilon=-U/2$
corresponding to the particle-hole symmetry point of the system.
However, in the FM case for the considered range of parameters,
no unitary limit of the conductance is achieved.
A completely different $D$-dependence of the conductance,
on the contrary, is seen for $\varepsilon=-U/3$.
Here, $G$ only slightly changes for the AFM configuration
in the range of values of the uniaxial anisotropy
parameter under consideration [open points in Fig.~\ref{Fig:9}(d)],
while it remains practically constant for the FM coupling
[open points in Fig.~\ref{Fig:9}(c)].
The origin of the observed behavior can be relatively easy
understood for $\varepsilon=-U/3$, where due to  the presence
of the effective dipolar exchange field the ground state doublet is split,
which leads to the suppression of linear conductance,
and for $\varepsilon=-U/2$ in the AFM case of the $J$-coupling
(see the above explanation for the antiparallel magnetic configuration).
On the other hand, for $\varepsilon=-U/2$ and the FM $J$-coupling
one could expect that in the absence of the dipolar exchange field,
the system should qualitatively behave somewhat
similarly as for the antiparallel magnetic configuration.
Closer analysis of the temperature dependence for $G_\text{P}$, Fig.~\ref{Fig:10}(a),
however, indicates that no Kondo effect arises at low temperatures,
and $G_\text{P}$ takes a relatively low value,
as compared to $G_\text{AP}$, in the temperature range of interest.
This can be attributed to the presence the effective \emph{quadrupolar} exchange field,
recently predicted to occur in large-spin ($S>1/2$) nanoscopic systems.~\cite{Misiorny_NaturePhys.9/2013}
A large-spin system subject to such a field
can acquire uniaxial magnetic anisotropy,
even though it was generically spin-isotropic,
and this effect is of pure spintronic origin due to the proximity of ferromagnetic electrodes.
As stemming from higher-order tunneling processes,
the quadrupolar exchange field is proportional to $\Gamma^2$,
and its effect is usually overpowered by the dipolar exchange field,
which is the first-order effect ($\propto\Gamma$).
Importantly, the quadrupolar field does not vanish
at the particle-hole symmetry point, as the dipolar field does,
where it can play an essential role especially in the case
of systems with no or small magnetic anisotropy.

In the situation under consideration,
see full points in Fig.~\ref{Fig:9}(c), one can see that for $D/T_\text{K}^0\lesssim10^{-2}$
the conductance takes a constant value,
which for small \emph{intrinsic} magnetic anisotropy
remains also independent of temperature, Fig.~\ref{Fig:10}(a).
In the light of the preceding discussion,
one can thus conclude that the transport properties of the MQD
in such a case are determined by uniaxial magnetic anisotropy
of spintronic origin (i.e., due to the quadrupolar exchange field),
so that the effect typical for the situation of $D \gg T_{\rm K}$ is observed.
Finally, we note that similarly as the dipolar exchange field,
also the quadrupolar field is absent for the antiparallel magnetic configuration of electrodes.

The corresponding dependence of the linear conductance on $D$
in both magnetic configurations is reflected in the TMR, which
is shown in Figs.~\ref{Fig:9}(e) and (f).
For ferromagnetic $J$ and $\varepsilon=-U/2$,
the TMR first decreases to reach a local minimum, then
increases to drop again and reach negative value.
For $\varepsilon=-U/3$, the TMR is generally negative
and depends rather weakly on $D$.
The dependence, however, changes completely in the case
of antiferromagnetic $J$. Now, the TMR becomes greatly enhanced
with decreasing $D$. This is related with the fact
that in the antiparallel configuration the conductance
should become fully suppressed in the limit $T\to 0$.
\cite{Misiorny_Phys.Rev.Lett.106/2011,Misiorny_Phys.Rev.B84/2011,Misiorny_Phys.Rev.B86/2012_UK}

\subsection{The effect of transverse magnetic field}

A characteristic, experimentally observed,
feature of a nanomagnet with an effective large spin $S$,
whose magnetic properties can be described by
the giant-spin Hamiltonian~(\ref{Eq:Ham_S}),
are oscillations of the tunnel splitting~$\Delta$ of the ground state
as a function of a magnetic field applied along
the system's hard anisotropy axis.~\cite{Wernsdorfer_Science284/1999,
Wernsdorfer_Phys.Rev.Lett.95/2005,Ramsey_NaturePhys.4/2008}
These oscillations are a manifestation of the quantum-mechanical
nature of the system under discussion, as they stem from
destructive interference between different tunneling paths.
\cite{Loss_Phys.Rev.Lett.69/1992,Delft_Phys.Rev.Lett.69/1992}
It was shown that the degeneracy of the ground state is
restored at some specific values $B_{x,\text{res}}^{(n)}$ of the field,
occurring  at the same interval $\Delta B_x=2\sqrt{2E(D+E)}$.
\cite{Garg_Europhys.Lett.22/1993,Villain_Eur.Phys.J.B17/2000,Kececioglu_Phys.Rev.B63/2001,
Bruno_Phys.Rev.Lett.96/2006}
The index $n\in \mathbb{N}$ labels the consecutive values of the field,
different from zero, for which $\Delta=0$, and as far as the
ground state splitting is considered $n$ cannot be larger
than the spin number of the system.
\footnote{In particular, for an integer $S$ one has $B_{x,\text{res}}^{(n)}=(2n+1)\sqrt{2E(D+E)}$,
whereas for a half-integer $S$, $B_{x,\text{res}}^{(n)}=2n\sqrt{2E(D+E)}$.
Note that for $n=0$ and an integer (half-integer) $S$ one
obtains $B_{x,\text{res}}^{(0)}=\sqrt{2E(D+E)}$ ($B_{x,\text{res}}^{(0)}=0$),
which is a consequence of the Kramers theorem.
For details see, e.g., Ref.~[\onlinecite{Gatteschi_book}]}
Importantly, although here we limit our discussion
to the ground state and the field applied along the hard ($x$) axis,
generally the degeneracy restoration can take place between any two,
repelling each other, states and also for the specific combinations
of the field components along the hard and easy ($z$) axes.
Such a point of the parameter space where this takes place
is commonly referred to as a \emph{`diabolical'} point.~\cite{Gatteschi_book}

\begin{figure}
\centering
\includegraphics[width=0.965\columnwidth]{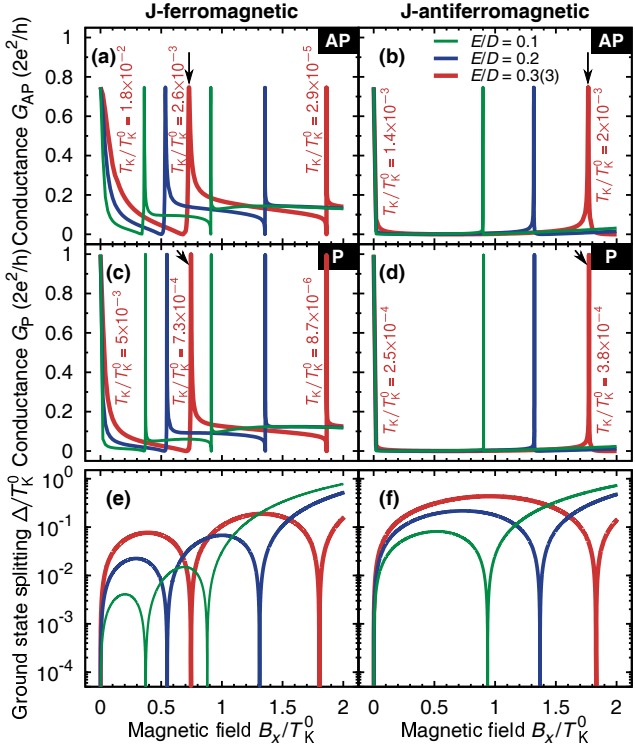}
\caption{(color online)
(a)-(d) Similar to Fig.~\ref{Fig:5}, but with the linear conductance $G$
at the particle-hole symmetry point ($\varepsilon=-U/2$)
plotted now as a function of magnetic field $B_x$
(i.e., applied parallel to the MQD's hard axis)
for several values of $E/D$ and $T/T_\text{K}^0=5\times10^{-10}$.
The indicated values of the Kondo temperature $T_\text{K}$
correspond the case of $E/D=1/3$, and has been derived from
the temperature dependence of conductance at respective fields
where the degeneracy of the ground state doublet is restored.
For specific magnitudes of these fields see the right panel of Fig.~\ref{Fig:14}.
\emph{Bottom panel} [(e)-(f)]:
oscillations of the ground states doublet splitting $\Delta$
due to the presence of transverse magnetic anisotropy for the ferromagnetic
(e) and antiferromagnetic (f) exchange interaction parameter $J$,
calculated for  an \emph{isolated} MQD.
Remaining parameters as in~Fig.~\ref{Fig:3}.
}
\label{Fig:11}
\end{figure}

Let us analyze such oscillations of the ground state doublet
in the case of the system under consideration, that is a MQD in the Kondo regime.
First and foremost, we recall that the total spin $S^t$ of
a MQD arises owing to the exchange interaction between
the spin of an electron occupying OL and the magnetic core spin,
cf. Hamiltonian~(\ref{Eq:Ham_MQD}), and only the latter
is represented by the giant-spin Hamiltonian~(\ref{Eq:Ham_S}).
For this reason, since the system is now described by more parameters than
only the anisotropy constants $D$ and $E$,
one should not generally expect that the degeneracy
will be restored at the constant interval $\Delta B_x$.

\begin{figure}[t]
\centering
\includegraphics[width=0.965\columnwidth]{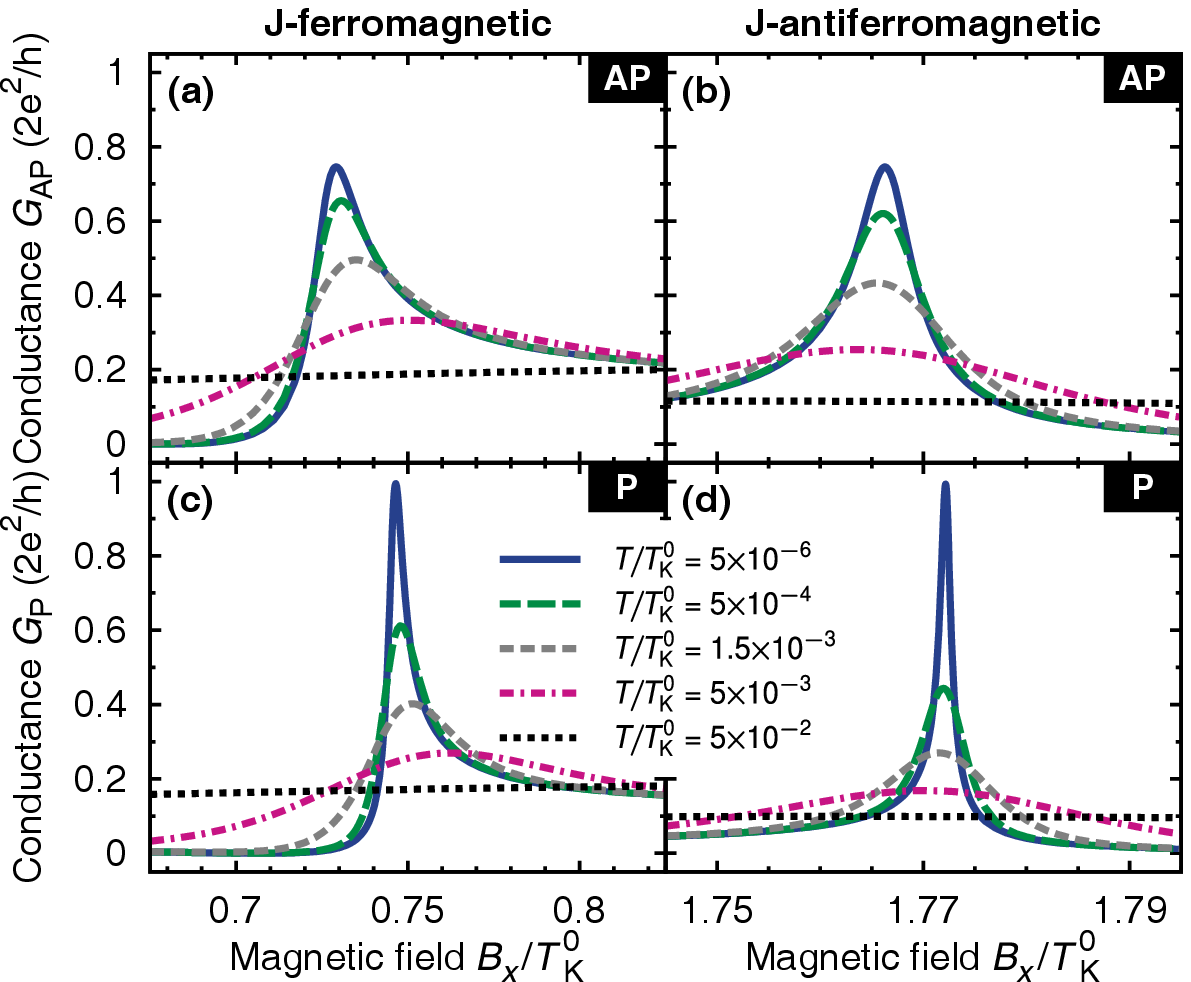}
\caption{(color online)
Temperature evolution of the Kondo resonance
around the first resonant field, $B_{x,\text{res}}^{(1)}$,
(marked with arrows in Fig.~\ref{Fig:11}) in the case of
the ferromagnetic (left column) and antiferromagnetic (right column)
exchange coupling for $E/D=1/3$. As in previous figures,
also two magnetic configurations of electrodes are considered:
antiparallel (top panel) and parallel (bottom panel).
Except temperature $T$, all other parameters are the same as in~Fig.~\ref{Fig:11}.
}
\label{Fig:12}
\vspace*{\baselineskip}
\includegraphics[width=0.965\columnwidth]{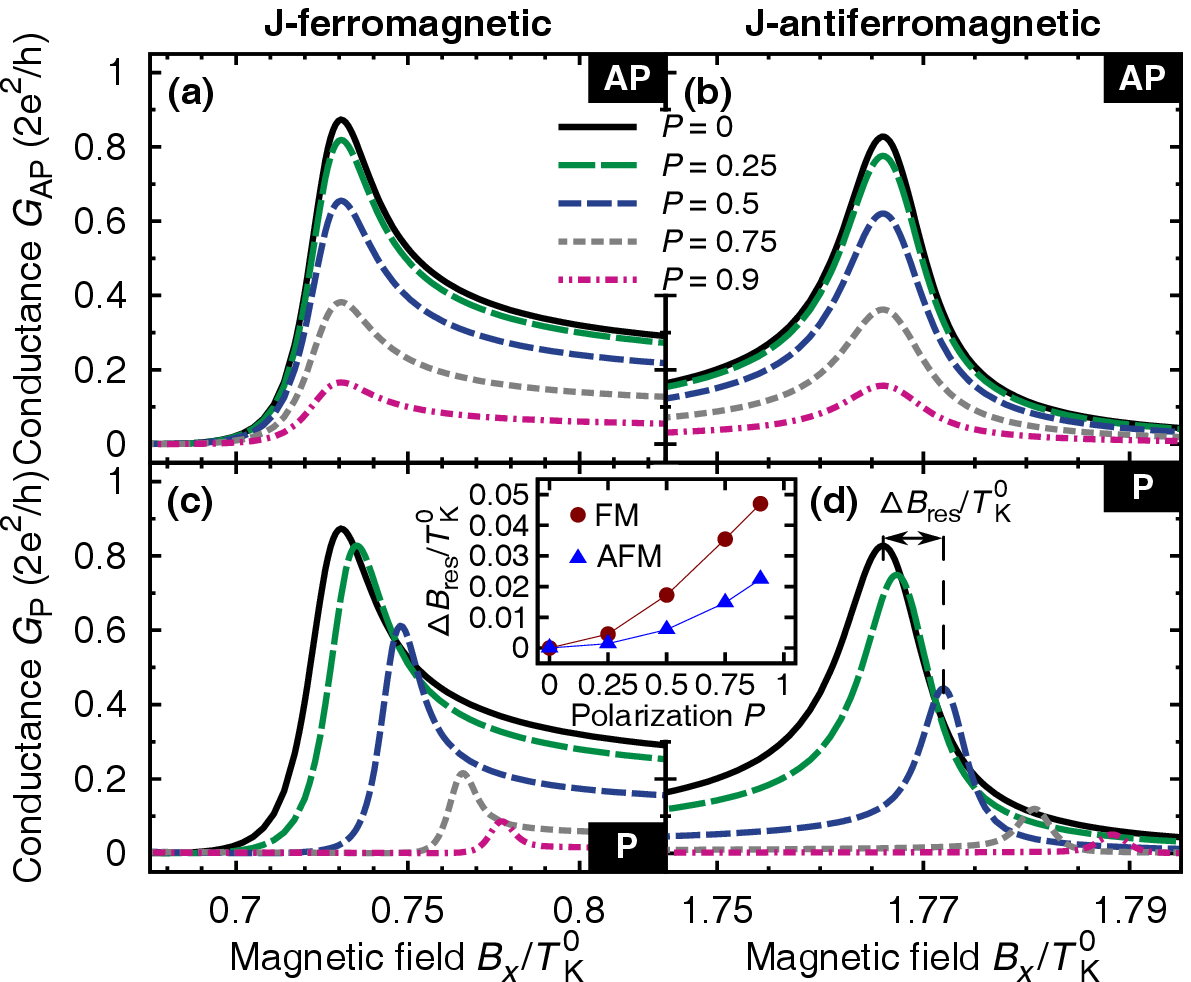}
\caption{(color online)
Analogous to Fig.~\ref{Fig:12}, but now
the dependence of the field-induced restoration
of the Kondo resonance on the spin polarization
coefficient $P$ of the leads is analyzed for $T/T_\text{K}^0=5\times10^{-4}$.
The inset presents the shift of the peak position $\Delta B_\text{res}$
as a function of $P$ in the parallel magnetic configuration,
measured with respect to the case of $P=0$ (solid line).
Remaining parameters as in Fig.~\ref{Fig:12}.
}
\label{Fig:13}
\vspace*{-\baselineskip}
\end{figure}

For a half-integer spin, as considered in this paper,
in the absence of an external magnetic field (and also the effective dipolar exchange field)
the ground state is twofold degenerate ($\Delta=0$), see Fig.~\ref{Fig:11}(e)-(f).
Then, as discussed above, if $E\neq0$, at sufficiently low temperatures,
$T<T_\text{K}$, one expects the Kondo effect to arise, Fig.~\ref{Fig:11}(a)-(d).
However, as soon as the field along the hard axis is applied the doublet
becomes split ($\Delta\neq0$) and  for $B_x\gtrsim T_\text{K}$ the
Kondo effect vanishes.
Since $T_\text{K}$ decreases as $E/D$ gets smaller,
one can easily see that the detrimental influence of the field
on the Kondo resonance will be more pronounced for systems
with weaker transverse magnetic anisotropy, compare thin
(small $E/D$) and bold (large $E/D$) lines in Fig.~\ref{Fig:11}(a)-(d).
Furthermore, as the magnitude of the field is increased,
whenever it approaches one of its values $B_{x,\text{res}}^{(n)}$
corresponding to the restoration of the degeneracy of the ground state doublet,
one observes that the Kondo resonance builds up again.
This process occurs in the field range around
$B_{x,\text{res}}^{(n)}$ whose energy scale is given by $T_\text{K}$, Fig.~\ref{Fig:11}(a)-(d).
In agreement with theoretical predictions about how
many times the degeneracy can be reinstated,~\cite{Gatteschi_book}
for the FM $J$-coupling ($S^t=5/2$) we observe two revivals of the Kondo effect,
whereas for the AFM $J$-coupling ($S^t=3/2$) only one.
Also, as expected, the maxima of conductance do not appear periodically.
In addition, one can notice that whereas for the FM exchange
coupling the width of the resonance becomes smaller for each next resonant field,
the opposite effect is observed in the AFM case,
compare the bold lines Fig.~\ref{Fig:11}(a)-(d), representing $E/D=1/3$,
for which values of $T_\text{K}$ at each resonant field have been provided.
Since with  lowering $E/D$ values of $T_\text{K}$ decrease,
Fig.~\ref{Fig:8}, this justifies an extremely low value of temperature $T$
used in calculations of Fig.~\ref{Fig:11}(a)-(d), which was to ensure
the occurrence of all possible resonances for given ratios $E/D$.
Finally, we note that the conductance maxima in Fig.~\ref{Fig:11}(a)-(d)
develop at somewhat different fields that one could expect from
calculations of the ground-state splitting $\Delta$ for an isolated MQD,
shown in Fig.~\ref{Fig:11}(e)-(f). This can be attributed to renormalization
of energy levels due to strong tunnel-coupling which, in turn,
leads to renormalization of the anisotropy parameters
$D$ and $E$,~\cite{Misiorny_Phys.Rev.B86/2012_UK,Oberg_NatureNanotechnol.9/2014}
whose values determine the resonant field.

\begin{figure*}[t]
\includegraphics[scale=0.7]{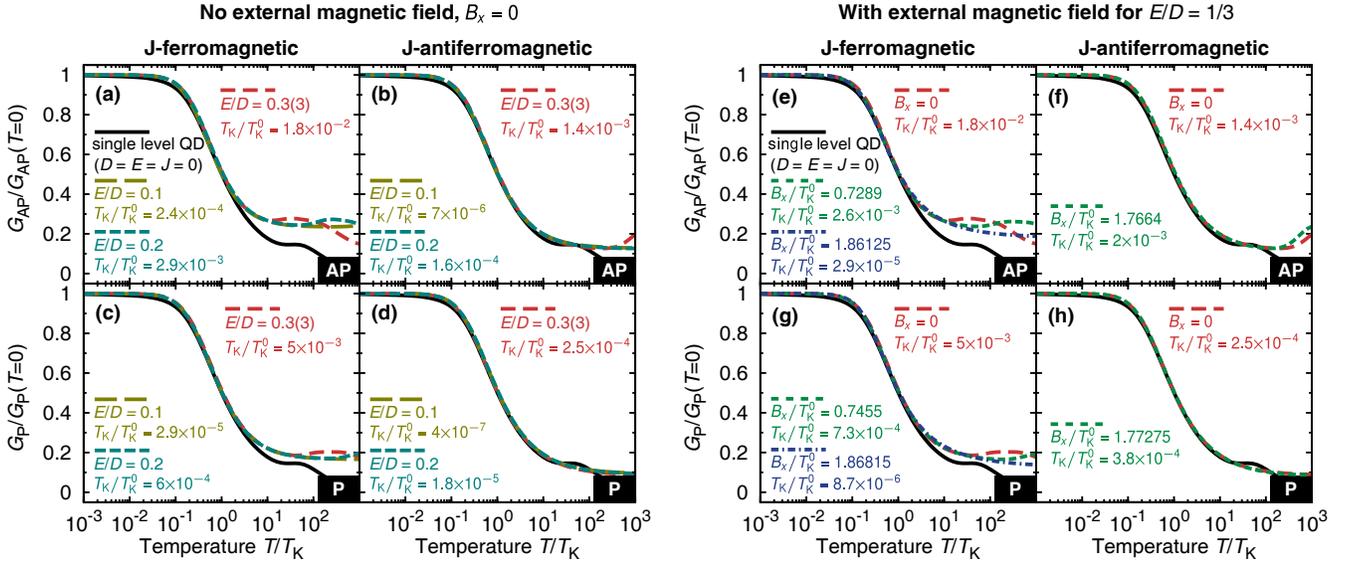}
\caption{(color online)
Universal features of the Kondo effect restored
by the presence of transverse magnetic anisotropy
at the particle-hole symmetry point ($\varepsilon=-U/2$).
\emph{Left panel} [(a)-(d)]:
Analogous to Figs.~\ref{Fig:6}(a)-(d),
but now the conductance is normalized to its value at $T=0$,
whereas the temperature $T$ is scaled with respect to
the Kondo temperature $T_\text{K}$ for a given $E$
[for specific values see the description of lines in each plot;
recall that $T_\text{K}^0/W\approx0.002$].
\emph{Right panel} [(e)-(h)]:
The scaling of conductance $G$ for $E/D=1/3$
[cf. bold lines in Figs.~\ref{Fig:11}(a)-(d)] is shown
for the values of magnetic field $B_x$ at which Kondo effect is restored.
In both panels solid lines are added to
facilitate the comparison with the case of a single
level quantum dot (QD),
whereas long-dashed lines in the right panel
allow for comparison with the case when the magnetic field is absent ($B_x=0$).
All other parameters as in Fig.~\ref{Fig:3}.
}
\label{Fig:14}
\end{figure*}

To get further insight into the properties of the Kondo effect restored
by means of the transverse magnetic field, we now focus our attention
on the peaks occurring for $E/D=1/3$ at the first resonant field,
$B_{x,\text{res}}^{(1)}$, in the case of different magnetic configurations
and sign of the exchange interaction parameter $J$.
These peaks are indicated in  Fig.~\ref{Fig:11}(a)-(d) with arrows.
First, in Fig.~\ref{Fig:12} we investigate the temperature
evolution of the conductance maximum developing at $B_{x,\text{res}}^{(1)}$.
It can be seen that, as expected, with the increase of temperature
the maximum becomes gradually smeared out, and eventually
the Kondo effect no longer shows up. From the width of the peak at low temperatures,
that is for which the peak has already reached its maximal available value,
we can qualitatively confirm that in general the Kondo temperatures are
lower for the parallel magnetic configuration.
Moreover, since now the field range under consideration
is limited to a vicinity of the resonant field, one can immediately notice
that for a given type of the $J$-coupling the exact values of $B_{x,\text{res}}^{(1)}$
differ for the antiparallel and parallel magnetic configuration.
In particular, in the parallel case the maximum occurs at a slightly larger value of the field.
Because, as mentioned above,  the value of $B_{x,\text{res}}^{(n)}$,
depends on the magnetic anisotropy, one can thus suspect this effect
may be related to the presence of the effective quadrupolar field in the parallel configuration.
Recall that we consider here the system at the particle-hole
symmetry point ($\varepsilon=-U/2$), so the dipolar exchange field is absent.

The effective quadrupolar exchange field is a spintronic effect,
which means that its magnitude depends on the spin polarization
$P$ and magnetic configuration of electrodes.~\cite{Misiorny_NaturePhys.9/2013}
Particularly, it grows as $P^2$ and gets switched off in the antiparallel configuration.
For these reasons, in order to check whether the shift of the conductance
maximum originates from the quadrupolar field, for a chosen temperature
in Fig.~\ref{Fig:13} we analyze how the position of the peak depends on $P$.
We find that while for the antiparallel magnetic configuration the peak,
albeit with a different height, always occurs at the same value of the field,
Fig.~\ref{Fig:13}(a)-(b), in the parallel configuration the maximum
moves towards larger fields as $P$ is increased, Fig.~\ref{Fig:13}(c)-(d),
and this effect is more pronounced for the FM $J$-coupling, see the inset in Fig.~\ref{Fig:13}.

Another interesting feature visible in the dependence of the linear conductance
on the transverse magnetic field is the asymmetry of restored Kondo resonances
with respect to the restoration field $B_{x,{\rm res}}^{(n)}$, see Figs. \ref{Fig:11}-\ref{Fig:13}.
This effect results directly from the asymmetry of corresponding matrix elements of the total spin
relevant for the spin-flip exchange processes, which are responsible
for the occurrence of the Kondo effect. \cite{Wegewijs_NewJ.Phys.9/2007}
The asymmetry of matrix elements gives rise to
the corresponding behavior of the Kondo peak as a function of $B_x$.

\subsection{Universal scaling}

Finally, we discuss the universal features of the Kondo resonance restored
by the presence of transverse magnetic anisotropy.
In particular, we analyze the normalized linear conductance $G/G(T=0)$
as a function of temperature scaled with respect to the Kondo temperature.
This allows us to check whether the temperature dependence
of the conductance follows that observed for a conventional single-level quantum dot.
For this purpose, we first consider the case when an external magnetic field
is absent ($B_x=0$), see the left panel of Fig.~\ref{Fig:14}.
Distinct dashed lines correspond there to different values of the transverse
magnetic anisotropy parameter $E$, while the solid line presents
the temperature dependence of $G$ for the case of a single-level quantum dot ($D=E=J=0$).
It should be emphasized that the Kondo temperature $T_\text{K}$
used for rescaling the temperature axis actually varies for each line,
and its specific values are given in the figure.
One can see that regardless of the type of the exchange coupling
the agreement between the conventional spin-1/2
and the pseudospin-1/2 Kondo effect discussed in this paper
is obtained  for $T/T_\text{K}\lesssim 1$. In the case of $T/T_{\rm K}>1$,
for the FM $J$-coupling the values of conductance for
the spin-anisotropic system can significantly exceed those
for the quantum dot [especially in the AP magnetic configuration,
see Fig.~\ref{Fig:14}(a)], whereas in the AFM case the universal
behavior of conductance is found up to the high temperature regime, $T/T_\text{K} > 1$.

The above analysis can be extended to the situation of an external
magnetic field applied along the MQD's hard axis.
In the right panel of Fig.~\ref{Fig:14} we show the
dependence 
of linear conductance on temperature 
in the case when the Kondo effect
arises owing to the field-induced oscillations of the ground
state doublet splitting. In particular, the dashed lines represent
the temperature dependence of the conductance at the maxima
appearing at some resonant fields in
Fig.~\ref{Fig:11}(a)-(d) for $E/D=1/3$ (bold lines).
We find in this case the same universal scaling
properties of the Kondo effect as those discussed above.


\section{Conclusions}


In this paper we analyzed the linear response transport properties
of a large-spin molecule strongly coupled to external ferromagnetic leads.
The main focus was on the role the transverse magnetic
anisotropy plays in formation of the Kondo effect.
The considerations were performed with the aid of the
full density-matrix numerical renormalization group method,
which allowed us to obtain accurate results for the studied system.
In particular, we analyzed the dependence
of the spectral functions on the orbital level position of the molecule,
the magnetic configuration of the device and the type of exchange
coupling between the magnetic core of the molecule and
its orbital level.

We showed that an additional finite transverse component of magnetic anisotropy
has a profound effect on transport characteristics of the system as it
can generally lead to the restoration of the Kondo resonance,
with the Kondo temperature depending now on the transverse anisotropy constant $E$.
Whereas  in the antiparallel configuration at sufficiently low temperature
the Kondo effect occurs as soon as the system enters
the local moment regime, $-U<\varepsilon<0$, in the parallel configuration the Kondo effect
is restored only at the particle-hole symmetry point ($\varepsilon=-U/2$), with considerably
smaller Kondo temperature.
Such a behavior is due to the presence of the effective dipolar exchange field
in the parallel configuration that splits the ground state doublet
and it vanishes only at that specific symmetry point.
In consequence, the influence of the transverse magnetic anisotropy
is most prominent at $\varepsilon=-U/2$ and it manifests especially
in the non-monotonic dependence of the tunnel magnetoresistance,
which for $E=0$ remains approximately constant.
Furthermore, the interplay of temperature and both the anisotropy parameters
was explored to establish the parameter space for which the Kondo effect can take place.

Finally, we also investigated the response of the molecule
to an external magnetic field applied along the system's hard axis,
expecting that the oscillations of the ground state splitting should
translate into periodic reoccurrence of the Kondo resonance.
We found that, unlike for large-spin nanomagnets, which can be described
by the giant-spin Hamiltonian, the resonant fields at which
the degeneracy restoration takes place do not appear
at the same interval depending only on the magnetic anisotropy parameters.
Interestingly, we showed that these fields hinge on the
magnetic configuration of electrodes and their spin polarization.
In particular, at the particle-hole symmetry point for the parallel
magnetic configuration we observed that with increasing
the spin polarization the Kondo resonances are reinstated
at slightly larger fields as compared to the antiparallel configuration,
where no similar dependence arises. We attribute this effect to the presence
of the effective quadrupolar exchange field,
recently proposed to exist in large-spin nanosystems.~\cite{Misiorny_NaturePhys.9/2013}


\acknowledgments
We acknowledge stimulating discussions with J.~Barna\'{s}, T.~Costi and M.~Wegewijs.
The use of the SPINLAB computational facility and the open access Budapest flexible DM-NRG
code~\cite{Legeza_DMNRGmanual,Toth_Phys.Rev.B78/2008}
(http://www.phy.bme.hu/\textasciitilde dmnrg/) is kindly acknowledged.
M.M. acknowledges the financial support from
the Alexander von Humboldt Foundation and from  the National Science
Center in Poland as the Project No. DEC-2012/04/A/ST3/00372.
I.W. acknowledges support from the National Science Center
in Poland as the Project No. DEC-2013/10/E/ST3/00213
and the EU grant No. CIG-303 689.



%

\end{document}